\documentclass[pdflatex,sn-aps]{sn-jnl}
\usepackage{csquotes}
\usepackage{fix-cm}
\usepackage{array}
\usepackage{eurosym}
\usepackage{titlesec}
\titlespacing*{\section}{0pt}{9pt}{5pt}
\titlespacing*{\subsection}{0pt}{9pt}{5pt}
\usepackage{graphicx} 
\usepackage{multirow}%
\usepackage{amsmath,amssymb,amsfonts}%
\usepackage{amsthm}%
\usepackage{mathrsfs}%
\usepackage{xcolor}%
\usepackage{textcomp}%
\usepackage{manyfoot}%
\usepackage{booktabs}%
\usepackage{algorithm}%
\usepackage{algorithmicx}%
\usepackage{algpseudocode}%
\usepackage[utf8]{inputenc}
\usepackage[normalem]{ulem}
\usepackage{listings}\usepackage{rotating} 
  \usepackage{tcolorbox}
\usepackage[
singlelinecheck=false 
]{caption}
\hypersetup{
  pdftitle={Hydrogen Mon Amour},
  pdfauthor={Christophe Grojean and Patrick Janot},
  pdfsubject={Hydrogen storage, natural hydrogen, and the electricity supply of the Future Circular Collider},
  pdfkeywords={FCC, CERN, hydrogen, natural hydrogen, energy storage, renewable electricity, sustainability}
}

\usepackage{etoolbox}   
\makeatletter
\patchcmd{\@maketitle}{\artauthors}{\centering{\artauthors}}{}{}
\makeatother

\begin{document}
\title{Hydrogen mon amour}

\author[1,2]{\fnm{Christophe} \sur{Grojean}}
\author*[3]{\fnm{Patrick} \sur{Janot}}\email{patrick.janot@cern.ch}

\affil[1]{\small \orgname{Deutsches Elektronen-Synchrotron DESY}, \orgaddress{\street{Notkestr. 85}, \postcode{22607} \city{Hamburg}, \country{Germany}}}

\affil[2]{\small \orgname{\orgdiv{Institut für Physik}, Humboldt-Universit\"at zu Berlin}, \orgaddress{\postcode{12489} \city{Berlin}, \country{Germany}}}

\affil[3]{\small \orgname{CERN}, \orgdiv{EP Department}, \orgaddress{\street{1 Esplanade des Particules}, \city{Meyrin}, \country{Switzerland}}}

\abstract
{
The FCC’s projected average electricity consumption is 1.3 TWh per year. If supplied entirely by nuclear power, this would represent about 10\% of the annual output of a large third-generation nuclear reactor, together with a corresponding share of its radioactive waste. By 2050, however, renewable sources could provide up to 80\% of this electricity on a substantially decarbonised European grid, supported by long-term power-purchase agreements and flexible operation. This could reduce the residual nuclear requirement, and its associated waste, by up to a factor of five.

This essay asks whether hydrogen could help bridge the remaining periods of low wind and solar generation. It examines production from renewable electricity, transport as hydrogen or ammonia, storage and conversion back into electricity, together with their efficiency losses and carbon footprints. It then considers the still uncertain prospect of naturally occurring hydrogen. None of these routes offers a ready-made solution. But the decades before the FCC begins operation leave time to develop one—and CERN could help accelerate that development by becoming a credible long-term customer.
}
\maketitle


\vfill
\noindent{\small\it \textbf{Foreword.}
This is the eighth essay in a series~\cite{janot2026essays}, since brought together in a single synthesis in Ref.~\cite{janot2026climatesciencefccthink}. The first
seven critically examined a number of claims about the environmental
impact of the FCC; the previous essay ended with a call for concrete,
constructive proposals. This one answers that call. Physicists should
identify problems, seek solutions, test them against
orders of magnitude and share them as widely as possible, thereby maximising their chances of being challenged, improved and, where appropriate,
implemented. What follows is a proposal for discussion, not a prediction or a settled programme. Its assumptions and figures are open to scrutiny and improvement.
If it proves sound, it has already ceased to belong to its authors.
The views expressed are those of the authors alone and do not represent
the position of CERN.}
\eject

\section{Electricity without radioactive waste?}
\label{sec:electricity}

The FCC-ee is often portrayed as an energy monster~\cite{janot2026monster}. Its projected average electricity consumption would be 1.3~TWh per year~\cite{benedikt2025fsr1}. If supplied entirely by nuclear power, this would represent about 10\% of the annual output of a large third-generation reactor, together with a corresponding share of its radioactive waste. By the time the FCC begins operation, however, renewable sources could provide up to 80\% of its electricity on a substantially decarbonised European grid. CERN has already set this transition in motion through contracts for 140~GWh of solar electricity per year~\cite{cern2024solar}. The collider's flexibility will also allow it to favour periods of abundance and reduce its power when renewable generation falls. Together, these measures could reduce both the residual nuclear requirement and the associated waste by as much as a factor of five.

Even so, nuclear power would probably still supply a significant minority share, perhaps 20--40\%, of the FCC's electricity needs. This is not a climate problem: nuclear power is among the lowest-carbon sources of electricity. But since its waste helped turn the FCC into a `monster', could that share be reduced still further? Hydrogen might offer a solution. But it would be neither the first choice nor even the second.

To see more clearly, let us follow the journey of energy, from the simplest option to the most uncertain. We shall begin by using it as it is generated, then try to carry it across a few hours with batteries. Hydrogen promises to carry it across days and distances, but every conversion takes its toll in energy and adds to its footprint. We shall measure that price on the scale of the FCC before asking which uses can truly justify it. Our investigation will then lead us beneath the soil of Lorraine, where a decisive question awaits: is natural hydrogen already a resource that can be exploited, or merely a promising discovery?

\section{Use first, store later}
\label{sec:storage}

The priority is straightforward: use renewable electricity when it is available~\cite{ecEnergyStorage}, because every conversion entails additional energy losses, equipment, and emissions. Converting electricity into hydrogen and then back into electricity will always be less efficient than using it directly.

\begin{quote}\itshape\normalsize
The most efficient conversion is the one we can do without.
\end{quote}

That still leaves the need to match generation with demand. Solar output peaks at midday; demand does not vanish when the sun sets. Wind turbines depend on the wind, and hydropower on river flows---not on the collider's schedule. As their share grows, electricity grids will alternate between periods of surplus and strain. Energy will therefore have to be shifted from the hours when it abounds to those when it is scarce.

Batteries are the second step. Their round-trip efficiency is around 85\%~\cite{doe2023liion,doe2022gridstorage}. They can respond within milliseconds~\cite{okafor2023inertia} to stabilise the grid, but can also absorb a surplus and release it over the following hours. Stationary sodium-ion batteries~\cite{iea2026sodium,iea2024batteries} could reduce reliance on lithium, nickel, and cobalt. Used electric-vehicle batteries~\cite{bobba2018secondlife} may still enjoy a second life, once their weight and volume become less restrictive. Batteries will not remove the need to generate electricity, but they will allow far less of it to go to waste.

\begin{quote}\itshape\normalsize
Batteries carry electricity across the hours.
\end{quote}

Their limit becomes apparent when electricity must be carried across days. Doubling the storage duration means roughly doubling the number of cells, even if they are used only a few times a year. Several days of reserve then become costly in both capital and materials. Reservoirs and pumped-storage hydropower can provide longer-lasting reserves, but their deployment depends heavily on geography, and droughts can restrict their use. Then come the dark doldrums: several cold, overcast, and windless days during which solar panels, wind turbines, and batteries can no longer meet demand.

\begin{quote}\itshape\normalsize
The third option must be able to carry electricity across the days.
\end{quote}

\section{Hydrogen carries electricity across the days}
\label{sec:hydrogen-storage}

To endure, energy must be able to wait. Hydrogen offers that possibility: it can be compressed, liquefied, or converted, then stored and transported. To recover the energy, hydrogen reacts with oxygen from the air, either electrochemically in a fuel cell or through combustion in a gas turbine coupled to a generator, thereby producing electricity on demand~\cite{doe2023bidirectional}.

Unlike a battery, extending the storage duration mainly requires a larger reservoir, without increasing the power rating of the production or reconversion equipment in the same proportion. This property becomes decisive for a reserve that is called upon only rarely but must remain available for several days~\cite{doe2023bidirectional}.

But first, that hydrogen has to be produced. Today, almost all global production relies on natural gas and coal. These fossil-based processes emit ${\rm CO}_2$ before the hydrogen is even transported or used. In 2025, low-emissions pathways still accounted for less than 1\% of global production~\cite{iea2025hydrogen}; they are expected to exceed that threshold for the first time in 2026~\cite{iea2026hydrogen}. Decarbonising our activities with hydrogen therefore requires us first to decarbonise its production.

The table below gives indicative carbon intensities for the main production pathways. Capture rates, methane leakage, equipment manufacture, and the energy consumed matter more than the colour on the label. These intensities refer to the production of one kilogram of hydrogen; they do not include the entire chain required to deliver it to France or Switzerland and convert it back into electricity. In particular, the EU requires emissions to be at least 70\% below the fossil-fuel benchmark for hydrogen to qualify as renewable or low-carbon~\cite{eu2023renewable,eu2025lowcarbon}.


\begin{table}[htbp]
\centering
\small
\begin{tabular}{>{\raggedright\arraybackslash}p{0.20\textwidth} >{\raggedright\arraybackslash}p{0.48\textwidth} >{\raggedright\arraybackslash}p{0.19\textwidth}}
\toprule
Common colour designation & Production pathway & Carbon intensity (kg${\rm CO}_2$e/kg ${\rm H}_2$) \\
\midrule
Black/brown & Gasification of coal or lignite & 22--26~\cite{iea2024ghg} \\
Grey & Natural gas reforming without carbon capture & 10--12~\cite{iea2024ghg} \\
Blue & Natural gas with ${\rm CO}_2$ capture and storage & 0.8--6~\cite{iea2023definitions} \\
Green (wind) & Wind-powered electrolysis & 0.4--0.8~\cite{iea2023definitions} \\
Green (solar) & Solar photovoltaic-powered electrolysis & 0.9--2.5~\cite{iea2023definitions} \\
Pink & Nuclear-powered electrolysis & 0.1--0.3~\cite{iea2023definitions} \\
White & Extraction of natural hydrogen & Must be measured case by case~\cite{brandt2023natural} \\
\bottomrule
\end{tabular}
\end{table}

To build our reserve of renewable electricity, the appropriate pathway is therefore water electrolysis, which splits water into hydrogen and oxygen. For every 100~kWh fed into the electrolyser, 60 to 70 are retained in the hydrogen~\cite{doe2026pem}. It must then be compressed, stored, and converted back into electricity. By the end of the chain, only about 30 usable kWh remain~\cite{doe2023bidirectional}. Technologies will improve; they will abolish neither the conversions nor the losses.

\section{The Sun's journey to CERN}
\label{sec:solar-journey}

The question that gave rise to this essay takes us far from CERN: to desert regions, rich in sunshine and close to the sea, where vast photovoltaic installations could power electrolysers. Even allowing for the manufacture of the panels and equipment, this green hydrogen would have a far smaller carbon footprint than the black, brown, or grey hydrogen produced from coal or gas. Wind power, whether used alone or alongside solar, could improve its carbon balance still further.

The difficulty reappears with transport. Hydrogen's extremely low density means that it must be compressed or liquefied, at the cost of additional energy and infrastructure. Another route is through ammonia, which is already transported by sea on a large scale. But ammonia must first be synthesised from hydrogen and nitrogen, transported, and perhaps cracked again on arrival before the hydrogen can be converted back into electricity. At every stage, energy is lost and another footprint is added: depending on the processes used, only 23--42\% of the initial energy is ultimately recovered~\cite{kojima2025roundtrip,giddey2017ammonia}. Burning ammonia directly, or using it in certain fuel cells, avoids the cracking stage---but not the losses, nitrogen oxides, or its toxicity.

Let us apply this chain to the FCC. Of its annual consumption of 1.3~TWh, nuclear power would still be expected to provide 0.26--0.52~TWh. Replacing all of that electricity with hydrogen produced far from CERN using solar power, then processed, transported, and converted back into electricity, would be particularly demanding. From the photovoltaic plant to the electricity returned to the grid, we conservatively assume an overall efficiency of 20\%. Between 1.3 and 2.6~TWh of photovoltaic electricity would therefore have to be generated. At 50--55~kWh per kilogram of hydrogen produced~\cite{jrc2025social}, this would require approximately 25,000--50,000 tonnes of hydrogen per year.

Extrapolating from the French photovoltaic plants under contract with CERN~\cite{cern2024solar}, this output would require a land area of roughly 8--16~km$^2$, equivalent to a square three to four kilometres on each side. In a sunnier region, the area could be smaller. The project is conceivable, but hardly immaterial: applying the average costs published by IRENA~\cite{irena2025costs}, the photovoltaic installations alone would require an investment of around \euro 1~billion, before even accounting for the electrolysers and the rest of the chain.

This calculation does not describe how the proposed system would operate; it establishes its maximum scale. Hydrogen would not have to replace nuclear power hour by hour, but would step in only after direct use, demand flexibility, and batteries had played their part. The actual requirement would therefore depend less on annual consumption than on the frequency and duration of the dark doldrums.

At a smaller scale, a study conducted for the FCC by the LAPLACE laboratory confirms this economic disadvantage. Replacing the emergency diesel generators with a hydrogen reserve capable of supplying 12~MW for 72~hours would increase the required investment almost ninefold, from \euro 4.8~million to \euro 42.5~million, without reducing operating costs~\cite{sapountzoglou2026fcc}.

\begin{quote}\itshape\normalsize
Renewable hydrogen is not an energy source. It is an expensive way of moving energy through time or across space.
\end{quote}

This chain would be absurd for shifting solar electricity from midday to evening every day. It may become useful during a few critical periods each year. Fire insurance remains valuable even when it is rarely called upon: the value of a reserve lies in being there when every other option falls short.

The solar pathway remains far better than producing hydrogen from coal or gas without ${\rm CO}_2$ capture. It is far less convincing when the aim is to turn it back into electricity at CERN, where low-carbon grid power is already available. On climate grounds alone, nuclear power would do better.

\begin{quote}\itshape\normalsize
The verdict does not condemn renewable hydrogen. It restores it to its proper role.
\end{quote}

That role begins where hydrogen is needed as a molecule: in the chemical industry, fertiliser production, refining, and perhaps, in future, the reduction of iron ore. It may extend to certain industrial heat requirements, maritime transport in the form of ammonia, and the production of synthetic fuels for aviation, which is difficult to electrify~\cite{ec2026refueleu}. For electricity storage, hydrogen must come after direct use, flexibility, and batteries---not before them.

The chain could also be shortened. Production in southern Europe would avoid some maritime transport, while hydrogen pipelines could connect producers, storage facilities, and consumers. The European Hydrogen Backbone envisages a network of approximately 53,000 kilometres by 2040, 60\% of it consisting of repurposed gas pipelines~\cite{ehb2022backbone}. That network is still only a project, and the carbon footprint of the hydrogen would continue to depend on how it was produced. Pipelines would nevertheless avoid systematically converting the molecule into ammonia, only to crack it again on arrival.

Yet all these solutions share the same drawback: the hydrogen must first be manufactured.

\section{What if Nature had already done the work?}
\label{sec:natural-hydrogen}

Hydrogen forms naturally underground through several geological reactions~\cite{erlach2026geological}. Oil, gas, and mining surveys have generally not looked for it: most available data come from samples collected for other purposes. The scarcity of documented occurrences therefore does not demonstrate that the resource itself is scarce; it may primarily reflect the absence of targeted exploration~\cite{erlach2026geological}.

In Lorraine, dissolved hydrogen has recently been detected in a former coal basin, with concentrations increasing\footnote{At Folschviller, hydrogen accounts for about 18\% of the mixture of gases dissolved in the water at a depth of 1,250 metres, but only 3~mg per litre of water~\cite{pironon2025lorraine}. For this measurement, a probe separates the gases from the water directly at the bottom of the borehole. La Fran\c{c}aise de l'\'Energie (FDE) is considering extending this principle to extract the hydrogen without bringing the water to the surface. The process currently under study does not involve hydraulic fracturing, but its ability to sustain an industrial flow rate remains to be demonstrated~\cite{fde2026lorraine}. In another borehole, at Pontpierre, hydrogen makes up 49.6\% of the gas mixture at a depth of 2,500 metres~\cite{fde2026lorraine}. Early simulations suggested a proportion of more than 90\% at 3,000 metres~\cite{bettayeb2023hydrogen}, although that projection remains to be verified.} with depth~\cite{bettayeb2023hydrogen,pironon2025lorraine}. Drilled to a depth of 3,655 metres in 2026, the PTH-2 borehole confirmed its presence across numerous geological intervals~\cite{fde2026lorraine}. Some extrapolations suggest that the Lorraine basin could contain as much as 46 million tonnes of hydrogen~\cite{bettayeb2023hydrogen}.

Forty-six million tonnes: a dizzying figure. It does not yet constitute a reserve, but let us use it as a working assumption. With an electricity conversion efficiency of 50\%, one kilogram of hydrogen, whose lower heating value is 33.3~kWh, would yield approximately 16.7~kWh of electricity~\cite{eurostat2023hydrogen}. Replacing the FCC's entire residual nuclear share would therefore require 16,000--32,000 tonnes a year. Over fifteen years, the total would be 0.25--0.5 million tonnes: about 1\% of the reported 46 million tonnes~\cite{bettayeb2023hydrogen}, and far less if the hydrogen were used only during the dark doldrums.

What ultimately matters, however, is the sustainable daily production rate. The Lorraine basin would have to sustain an output of 45--90 tonnes of hydrogen a day, whereas the PTH-2 borehole has yet to demonstrate an industrial flow rate of anything like that magnitude~\cite{fde2026lorraine}. A concentration is not a flow rate, and a discovery is not yet an industry. We need to determine how the hydrogen moves, how quickly it forms, whether the reservoir replenishes itself, which gases accompany it, how much energy its separation will require, and what fraction will escape. The integrity of the wells and the protection of the groundwater resources they pass through will also have to be guaranteed. Finally, the carbon intensity of the entire chain, including drilling, will have to be measured.

\begin{quote}\itshape\normalsize
Natural hydrogen will have to earn its low-carbon credentials with hard numbers.
\end{quote}

A preliminary assessment nevertheless offers a first indication of the order of magnitude. For a gas containing 85\% hydrogen and little methane, extraction and purification would emit approximately 0.4~kg${\rm CO}_2$e per kilogram of ${\rm H}_2$~\cite{brandt2023natural}. Assuming a conversion efficiency of 50\%, this footprint alone would amount to roughly 25~g${\rm CO}_2$e per kilowatt-hour of electricity produced~\cite{brandt2023natural,eurostat2023hydrogen}, before accounting for the conversion plant and grid injection.

Natural hydrogen, known as `white' in the industry's colour code, could be comparable to the best wind-powered green hydrogen, two to six times less carbon-intensive than solar-powered green hydrogen, and twenty-five to thirty times less carbon-intensive than grey hydrogen. The presence and leakage of methane, the indirect atmospheric effects of hydrogen leakage~\cite{sand2023gwp}, or purification powered by carbon-intensive energy could, however, substantially worsen this balance. These figures illustrate what a favourable reservoir might make possible, rather than providing an assessment of Lorraine itself.

This `white gold' could therefore be among the lowest-carbon forms of hydrogen available to us. More importantly, it would no longer be merely an energy carrier: it would become a primary resource, produced by nature without the losses inherent in electrolysis.

Would this be oil all over again, under another name? The analogy is tempting. It usefully reminds us that no form of extraction is without consequences---but that is where it ends. The climate was warmed by the carbon released from fossil resources. Natural hydrogen contains no carbon to burn.

Nor would its extraction be without consequences: drilling, water use, land take, leakage, associated gases, and effects on ecosystems would all have to be measured. Batteries, grids, solar panels, and wind turbines also require resources. The question, therefore, is not whether to extract at all, but what to extract, in what quantities, for which uses, and at what environmental cost.

\section{From promise to proof}
\label{sec:proof}

A discovery does not become an industry on the strength of promises alone. The first facilities must be financed before their market exists; a customer willing to make a long-term commitment can then reduce uncertainty and unlock investment. CERN has already played this role through the photovoltaic contracts it signed in 2024~\cite{cern2024solar}. If natural hydrogen were to fulfil its promise, a similar commitment could help bring the industry into being.

CERN would neither drill in Lorraine nor fill storage tanks at Meyrin: it would simply enter into a long-term purchase agreement with a producer of electricity generated from natural hydrogen. The hydrogen would be extracted and converted close to its source; the electricity would be fed into the grid when renewables and batteries were insufficient. The contract would impose verifiable criteria for production rates, leakage, cost, and carbon intensity.

The timing could be ideal. First revealed in Lorraine in 2023~\cite{bettayeb2023hydrogen}, natural hydrogen must now pass the test of its first production trials, planned for 2028--2029~\cite{fde2026lorraine}. Success would leave almost twenty years to confirm the flow rates, build an industry, and supply the FCC in 2048, just in time~\cite{cerncourier2023century}.

\begin{quote}\itshape\normalsize
CERN would become a launch customer.
\end{quote}

If white hydrogen proved scarce, using it to generate electricity would probably not be its best use. It should first replace fossil-derived hydrogen, whether black, brown or grey, in the chemical industry, fertiliser production, and steelmaking. A tiny fraction, called upon only during the dark doldrums, could nevertheless secure the FCC's electricity supply and further reduce the residual nuclear share---and, with it, the associated radioactive waste.

Natural hydrogen is not yet the `white gold' of the twenty-first century. That will depend on its abundance, accessibility, cost and carbon footprint. Above all, it must pass the test that separates a promise from a resource: it must emerge sustainably from the ground, at a measurable flow rate---and not merely in press releases.

Lorraine may be an exception. It may also be the first answer we have found because someone finally asked the question. How many other sources lie hidden beneath our feet, concealed less by their rarity than by our indifference?

\begin{quote}\itshape\normalsize
Hydrogen had been there for millennia. It was simply absent from the questions we asked.
\end{quote}

Natural hydrogen would itself be only one step. Today, we expend energy to manufacture hydrogen. Tomorrow, perhaps, we shall harness the hydrogen that the Earth produces without us. Later still, current research may enable us to master the fusion of two hydrogen isotopes, deuterium and tritium, and release on Earth the energy that makes the stars shine~\cite{iterFusion,eurofusionDemo,ifmifDones,w7x}.

From chemistry to geology, and then from geology to the heart of matter, hydrogen would accompany every stage in our search for energy. For now, natural hydrogen remains an alluring promise. Fusion is already a fascinating scientific adventure; its industrial and commercial exploitation remains a distant prospect. In both cases, the same rule applies: we must move from hope to experiment, and from experiment to numbers.

\begin{quote}\itshape\normalsize
In love as in science, promises are not enough. Evidence is required.
\end{quote}

\subsection*{Credits}

Original screenplay: Christophe Grojean and Patrick Janot\\
Inspired by a question from Daniel Treille: solar hydrogen in Oman~\cite{cleanenergywire2025oman}\\
Lorraine location scouting: G\'eraldine Servant\\
Written, directed, and edited by Patrick Janot\\
Script doctor and scientific review: Christophe Grojean\\
English subtitles: Guy Wilkinson\\
Produced with CERN's (decarbonised) energy.

\vfill\eject
\subsection*{Disclaimer} The French version of this essay was originally prepared by the authors for a French-speaking readership. Artificial-intelligence language tools assisted both in proofreading the French text and in translating it into English. The translation was reviewed by several native speakers of English. The authors remain fully responsible for the final text.

\bibliography{hydrogen_mon_amour}

\end{document}